\documentclass[10pt, conference, letterpaper]{IEEEtran}
\IEEEoverridecommandlockouts
\usepackage[draft, bookmarks=false]{hyperref}
\usepackage{cite}
\usepackage{amsmath,amssymb,amsfonts, dsfont, bm}
\usepackage{algorithmic}
\usepackage{graphicx}
\usepackage{multirow}
\usepackage{subcaption}  
\usepackage{textcomp}
\usepackage{xcolor}
\usepackage{booktabs}
\usepackage{newtxtext,newtxmath}
\usepackage[english]{babel}
\usepackage{soul}

\def\BibTeX{{\rm B\kern-.05em{\sc i\kern-.025em b}\kern-.08em
    T\kern-.1667em\lower.7ex\hbox{E}\kern-.125emX}}
\begin{document}

\title{Large Spectrum Models (LSMs): Decoder-Only Transformer-Powered Spectrum Activity Forecasting via Tokenized RF Data}


\author{\IEEEauthorblockN{Mohammad Mosiur Lunar $\;$ Mehmet C. Vuran
		}
	\IEEEauthorblockA{
		Cyber-Physical Networking Lab, School of Computing; University of Nebraska-Lincoln, Lincoln, NE, USA, \\
		Email:  \{mlunar2, mcv\}@unl.edu \vspace{-1.5em}}
}

\maketitle


\begin{abstract}
Dynamic spectrum access (DSA) has become a key pillar of next-generation wireless systems to address the spectrum scarcity due to the rapid growth of connected devices. Accurate short-term spectrum forecasting is critical for DSA, where data-driven approaches have proven most effective. Recent advances in and widespread adoption of large language model (LLM) architectures present new opportunities for spectrum prediction. In this paper, foundational large spectrum models (LSMs) are presented. A novel RF tokenizer is introduced to convert raw IQ measurements into token sequences by mapping each power-spectral density value to a fixed vocabulary along with embedding gain, frequency, FFT bin, and timestamp information. Five established open-source LLM architectures (Gemma-2B, GPT-2, LLaMA-7B, Mistral-7B and Phi-1) are trained on this tokenized spectrum data for the task of spectrum forecasting, yielding LSMs. To leverage the scaling gains of LSMs, a fully automated outdoor wireless testbed is employed to collect over $22$ TB of raw spectrum data across $33$ sub-GHz frequency bands, yielding $8.4$ B tokens in total. Across all 33 bands, the best model (LSM-Mistral) achieves a root-mean-square error of $3.25$ dB and $97\%$ of predictions have a mean absolute error below $5$ dB. Generalization of LSMs is illustrated by fine-tuning the models on data collected in different locations, where RMSE is maintained below $3.7$ dB. These results demonstrate that the widespread decoder-only transformer architectures can serve as effective predictors of spectrum activity. Moreover, the comprehensive LSM framework with tokenized RF data has been shown to scale to locations where large amounts of data may not be available. The dataset, tokenizer, and trained models are released on GitHub and Hugging Face (\url{https://lsm.unl.edu}, \url{https://github.com/UNL-CPN-Lab}).

\end{abstract}

\begin{IEEEkeywords}
Spectrum Sensing, Dynamic Spectrum Sharing, LSM, Tokenized RF Data, LLM
\end{IEEEkeywords}

\section{Introduction}
\label{sec:intro}
\vspace{-0.05in}

The rapid growth of connected devices and data-driven applications has placed unprecedented pressure on the wireless spectrum. By 2030, the number of Internet of Things (IoT) devices is expected to reach $40$ billion, generating massive volumes of wireless traffic~\cite{IoTA24}. Although higher-frequency solutions such as mmWave and terahertz bands are being explored for $5$G and beyond~\cite{Wang6G}, their limited range and high computational demands restrict widespread deployment. Consequently, efficient utilization of existing spectrum resources has become increasingly critical.

\begin{figure}[t!]
	\centering
		\includegraphics[width=0.46\textwidth]{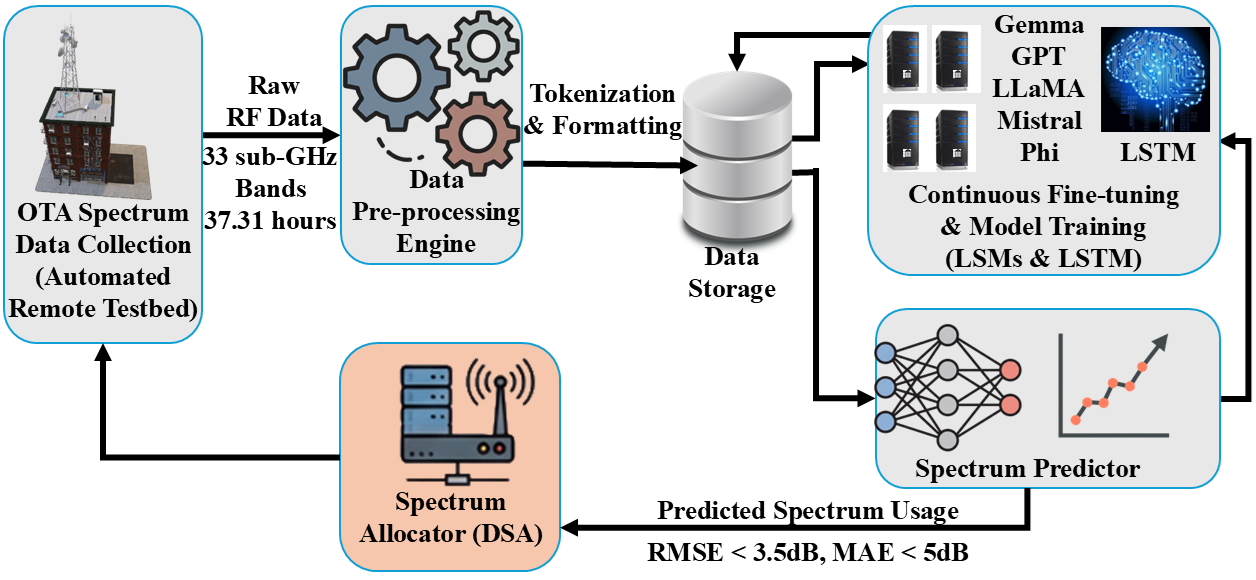}
		\vspace{-0.05in}
	\caption{\it Data-driven spectrum prediction system.}\label{fig:example_arch_lsm}
	\vspace{-0.25in}
\end{figure}

Dynamic Spectrum Access (DSA) has emerged as a promising approach to improve spectrum utilization by enabling opportunistic access to underutilized frequency bands~\cite{Akyildiz2006NeXt}. For example, the U.S. Citizens Broadband Radio Service (CBRS) employs a three-tier spectrum sharing framework coordinated by Spectrum Access Systems (SAS). Advances in software-defined radio (SDR) technology have made wideband spectrum sensing feasible~\cite{Manco22, Smith23}, producing large volumes of spectrum measurement data. However, effective DSA requires not only sensing current spectrum occupancy but also predicting future spectrum usage, making spectrum forecasting a key enabling capability. Accurate short-term spectrum forecasting could complement DSA systems, such as CBRS, by enabling more proactive channel selection and interference avoidance. More broadly, future 6G networks are expected to incorporate AI-driven spectrum intelligence~\cite{Cao24}, where predictive models may support data-driven and proactive spectrum management. An overview of the proposed data-driven spectrum prediction framework is illustrated in Fig.~\ref{fig:example_arch_lsm}.

Recently, a few spectrum prediction techniques have been explored with data-driven approaches, demonstrating superior performance in highly dynamic environments~\cite{Zhao19, Cullen23}. Recent progress in deep learning has further accelerated this trend, motivating the exploration of more expressive model architectures for spectrum forecasting tasks.

Large language models (LLMs) have recently demonstrated remarkable generalization capabilities across diverse domains. Their ability to model long-range dependencies and complex temporal patterns makes them appealing candidates for spectrum prediction. However, applying LLMs to spectrum forecasting presents several challenges. First, large-scale, high-quality spectrum datasets are required. Furthermore, LLMs require tokenized datasets, which are ideal for textual or linguistic data but may not naturally align with spectrum data. Finally, the resulting models must achieve sufficient accuracy to be useful for real-world spectrum management.

In this paper, we address these challenges by introducing foundational \textbf{Large Spectrum Models (LSMs)} for spectrum usage prediction. The key contributions of this paper are summarized as follows:

\begin{itemize}
 \item An novel RF tokenization algorithm is developed to represent the metadata and activity of the spectrum. The RF tokenizer converts raw spectrogram data into discrete tokens and subsequently returns the actual values during the evaluation phase.
 \item A broadband integrated geographical radio environment dataset (BIG-RED) has been constructed, containing spectrum sensing data collected from $33$ distinct sub-GHz frequency bands. BIG-RED was collected from rooftop and small cell sites at Nebraska Experimental Testbed of Things (NEXTT), a fully automated city-scale outdoor testbed~\cite{Zhao21Adhoc,OneLNK21}.
 \item The tokenized dataset is used to train five flavors of LSMs based on scaled-down versions of five open-source LLM architectures: Gemma-2B~\cite{Gemma24}, GPT-2~\cite{GPT19}, LLaMA-7B~\cite{LLaMA23}, Mistral-7B~\cite{Mistral23}, and Phi-1~\cite{Phi23}.  
 \item A comprehensive evaluation of the LSMs are provided, including a direct comparison with a non-LLM baseline based on time-series models, which has been widely used for spectrum prediction. In addition, LSM generalization is evaluated by fine-tuning the model on a separate geographical location. To the best of our knowledge, this is the \textbf{first end-to-end large sequence modeling framework} for the prediction of spectrum utilization.
 \item To support reproducible research, we release the spectrum dataset, tokenizer, trained models, and all associated code for data preprocessing, model development, training, validation, and testing at \url{https://lsm.unl.edu} and \url{https://github.com/UNL-CPN-Lab}.
\end{itemize}

The remainder of the paper is organized as follows: In Section~\ref{sec:related_lsm}, related work is reviewed. In Section~\ref{sec:data_and_problem_lsm}, the data collection strategy and the problem formalization are presented. In Section~\ref{sec:methodology_and_architecture_lsm}, the data processing methodology and the architecture of LSM are described. The evaluation results are presented in Section~\ref{sec:evaluation}, including both base variants and fine-tuned variants. Finally, we conclude the paper in Section~\ref{sec:lsm_conclusion}.

\vspace{-0.05in}
\section{Related Work}

\label{sec:related_lsm}

The influence of deep learning on wireless communication has grown rapidly~\cite{Abd24, Matsumine24}. From channel estimation~\cite{Guo24, Gao24} to beamforming~\cite{Wang25, Kang24}, it has enabled major advances across the field. In spectrum sensing, deep learning has been applied to signal classification~\cite{Shin25}, detection~\cite{Kumar24}, and prediction~\cite{El24}. Techniques such as real-valued signal analysis for joint signal detection and classification~\cite{Subedi24LCN} and complex-valued neural networks for signal segmentation~\cite{Shin25} have been used, while convolutional neural networks (CNNs) and recurrent neural networks (RNNs)~\cite{Kumar24} support detection tasks. Moreover, cognitive radio-based approaches~\cite{El24} explore spectrum prediction.

To train deep learning models, both real-world datasets~\cite{Doke24} and synthetic or laboratory-generated data~\cite{Uvaydov21} have been employed  for wireless communication applications. Depending on the specific objectives and design considerations, a diverse range of deep learning architectures have been adopted. Broadly, these architectures can be categorized into three: CNN-, RNN-, and transformer-based architectures, each tailored to address distinct characteristics of wireless signal processing tasks.

\begin{figure*}[t!]
	\centering
	\subfloat[Antenna array]{
		\includegraphics[height=1.5in]{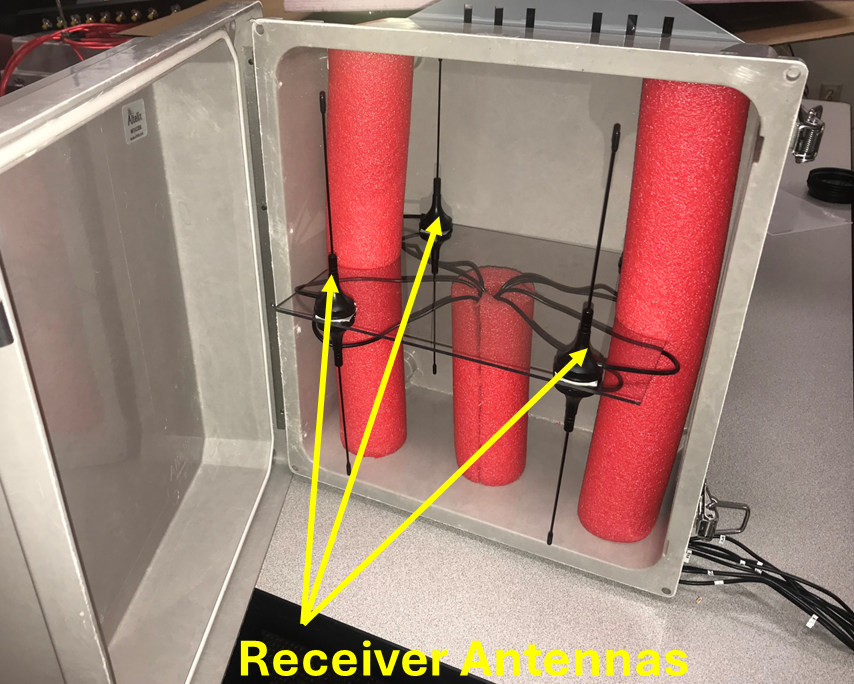}\label{fig:antenna_array}
	}
	\subfloat[Rooftop antenna]{
		\includegraphics[height=1.5in]{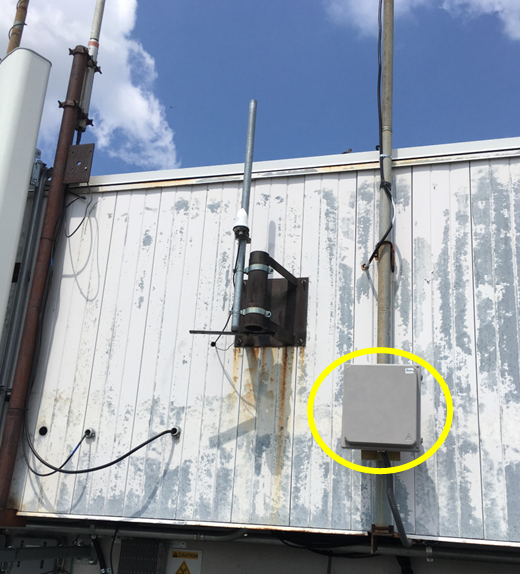}\label{fig:ofh_rooftop}
	}
	\subfloat[USRP N310]{
		\includegraphics[height=1.5in]{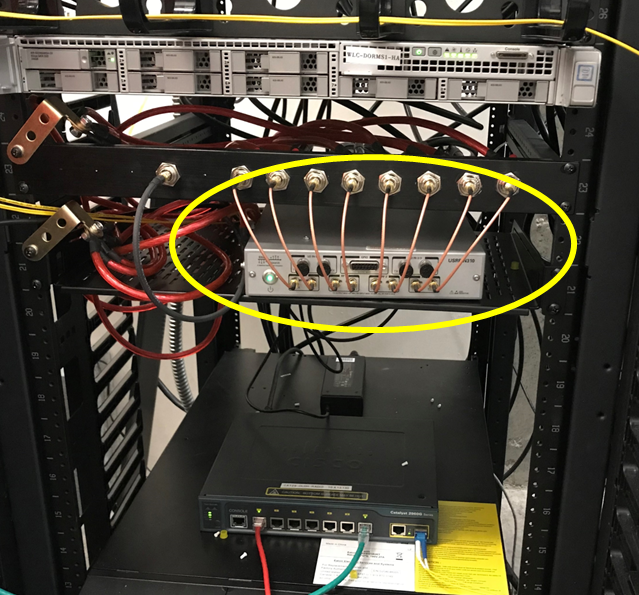}\label{fig:usrp}
	}
    \subfloat[Edge node]{
		\includegraphics[height=1.5in]{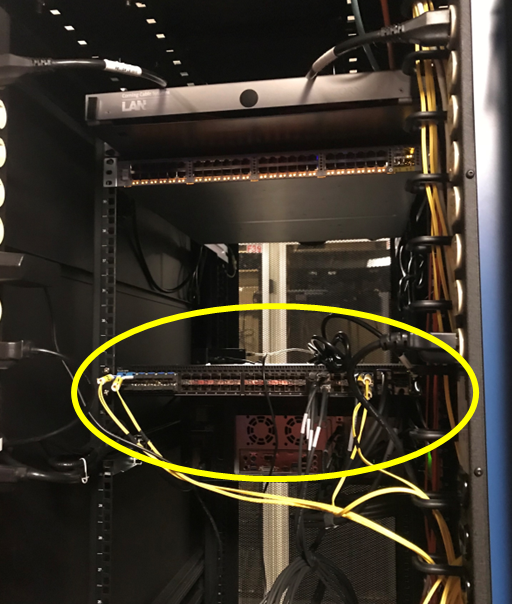}\label{fig:edge}
	}

	\vspace{-0.05in}
	\caption{\it Data collection setup at the core site~\cite{Zhao21Adhoc}.}\label{fig:data_collection}
	\vspace{-0.25in}
\end{figure*}

A 1D-CNN-based method for predicting wideband spectrum occupancy from narrowband observations is presented in~\cite{Uvaydov21}. A hybrid architecture that combines CNN and bidirectional LSTM (Bi-LSTM) networks for spectrum classification is proposed in~\cite{xing2022spectrum}, where features extracted by CNN layers are fused with raw input data to improve classification accuracy. A joint 2D-CNN and 1D-LSTM framework for signal classification is described in~\cite{Lees19}, and this architecture is further extended for signal detection and classification tasks in~\cite{Zhang21}. A 1D-LSTM-based model is developed in~\cite{Perenda21} for signal classification tasks. Additionally, a spectrum prediction technique based on differential minimization is introduced in~\cite{Li20}.

Recent research efforts have begun to explore transformer-based architectures for spectrum sensing tasks. A custom multi-head self-attention transformer designed for detecting activity in wideband spectrum environments is presented in~\cite{Zhang24}, where the model, trained on simulated datasets, outperforms conventional CNN-based approaches. A hybrid architecture that integrates transformer mechanisms with LSTM blocks in the encoder is proposed in~\cite{Guangliang25}, demonstrating improved performance over traditional LSTM-based models on laboratory-generated data. While these custom transformer designs show promising results, they often require a significant task-specific architectural tuning and development effort. Importantly, the application of general-purpose transformer-based LLMs to real-world spectrum sensing datasets remains largely unexplored, highlighting a compelling direction for future research.

In this work, we aim to bridge this research gap by developing and evaluating five LSM architectures derived from well-established LLM variants. These models are trained and validated using a real-world spectrum sensing dataset. To the best of our knowledge, this represents the first systematic effort to leverage LLM-based architectures for spectrum sensing on real-world data.
\vspace{-0.05in}
\section{Data Collection and Problem Formalization}
\label{sec:data_and_problem_lsm}

In this section, we lay the foundation for this work by describing the data acquisition methodology in detail. In addition, we formally define the problem and its scope.

\vspace{-0.1in}
\subsection{Spectrum Data Collection}
\label{subsec:datacol_lsm}

The broadband integrated geographical radio environment dataset (BIG-RED) is collected using the Nebraska Experimental Testbed of Things (NEXTT), a city-wide outdoor testbed~\cite{Zhao21Adhoc,OneLNK21}. NEXTT is equipped with USRP N310~\cite{ettus2025usrp_n310} SDR devices, connected through a fiber-optic backhaul network that enables seamless data collection from each site. NEXTT includes multiple SDR units deployed throughout the city. For this work, we select two sites: a core site and an auxiliary site. 

The core site is on the rooftop of a $12$-story building, representing a macro cell. In Figs.~\ref{fig:data_collection}, we show the equipment used in the core site. The UHF antenna array connected to the USRP N310 is shown in Fig.~\ref{fig:antenna_array}. The enclosed receiver and transmitter antenna arrays are mounted on the rooftop of the test site, as depicted in Fig.~\ref{fig:ofh_rooftop}. All antennas are connected to the USRP N310, located in the penthouse (Fig.~\ref{fig:usrp}), where an edge node is also installed to collect data and support real-time processing for the USRP (Fig.~\ref{fig:edge}). This edge node provides full remote access capability, enabling experimenters to conduct tests in a ``schedule-and-forget" manner.

\begin{figure}[t!]
	\centering
		\includegraphics[width=0.4\textwidth]{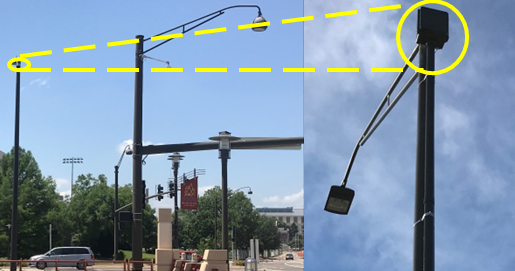}
		\vspace{-0.05in}
	\caption{\it Antenna array placement at the auxiliary site.}\label{fig:avtl_site}
	\vspace{-0.25in}
\end{figure}

The auxiliary site is located at a traffic intersection and models a small cell site (Fig.~\ref{fig:avtl_site}). This site is used to evaluate the generalization capabilities of LSMs using a fraction of the data compared to the core site for fine-tuning. The auxiliary site features an antenna array mounted on a street light pole, with a USRP N310 housed in an enclosed rack within a traffic signal cabinet. Similar to the core site, the USRP N310 is connected to the fiber-optic network for seamless data transfer. The distance between the core site and the auxiliary site is approximately $915\mathrm{m}$ and the height of the antenna array at auxiliary site is $3\mathrm{m}$ (around $42$m shorter than the core site).

Using this remotely accessible citywide testbed, we collected a total of $134{,}332$ seconds (equivalent to $37.31$ hours) of sub-GHz raw spectrum data from June 2023 to September 2023. Each spectrum sample spans a duration of $1s$ and covers a bandwidth of $20\mathrm{MHz}$. This results in a total of $2.69$ million MHz$\cdot$sec, amounting to $20.97$,TB of raw spectrum sensing data collected for this study. 

A total of $33$ distinct sub-GHz frequency bands between $[54\:\mathrm{MHz} $ - $990\:\mathrm{MHz}]$ are used for data collection. Each spectrum sample spans a bandwidth of $20\:\mathrm{MHz}$. The wideband UHF antenna installed with the USRP demonstrated excellent signal reception performance across all sub-GHz bands during laboratory experiments. After each sample is collected, the testbed stores the sample along with its frequency, bandwidth, collection time, and spectrum receiver gain on a storage server for future use. The testbed then switches the center frequency and collects samples sequentially.

\vspace{-0.05in}
\subsection{Problem Formulation}

\label{subsec:prob_des_lsm}

As mentioned in the previous section, at each site we collect IQ data along with their corresponding frequency, gain, and timestamp. Consider a measurement site $ST$, with a receiver gain of $G$ and a bandwidth of $B$. Let
$\{f_i\}_{i=0}^{M-1} = \{f_0, f_1, \dots, f_{M-1}\}$
denote the set of $M$ center frequencies under consideration. For each frequency, $f_i$, suppose that $R_i$ spectrum records are collected. This collection is denoted as $\mathcal{X}^{(i)} \triangleq \{\mathbf{x}_r^{(i)}\}_{r=0}^{R_i-1}, i=0,\dots,M-1$. The total number of spectrum records across all frequencies is $R \triangleq \sum_{i=0}^{M-1} R_i$.


Each spectrum record $\mathbf{x}_r^{(i)}\in \mathbb{C}^{N}$ is a complex vector of length $N=d\cdot B$, where $d$ denotes the acquisition duration of a single record. Then, the spectrum record is represented as 
$\mathbf{x}_r^{(i)}=[x_{r,0}^{(i)}, x_{r,1}^{(i)}, \cdots , x_{r,N-1}^{(i)}]^{\mathsf T}$. ,


Each element of this vector, $x_{r,n}^{(i)}$ is the $n$-th IQ sample within record $r$ at frequency $f_i$: $x_{r,n}^{(i)} = a_{r,n}^{(i)} + j\,b_{r,n}^{(i)}$,
where $a_{r,n}^{(i)}$ and $b_{r,n}^{(i)}$ are the in-phase and quadrature components, respectively.


We now have a total of $R$ records across $M$ frequency bands, each represented by a vector of raw IQ data of size $N$. Applying the short-time Fourier transform (STFT) to each spectrum record results in a $2D$ matrix representation of power spectrum density (PSD) or a spectrogram of the form
$\mathbf{S}_r^{(i)} \in \mathbb{R}^{n_s \times n_t}$,
where $n_s$ is the number of frequency bins and $n_t$ is the number of time slices. Its entries are denoted by $\mathbf{S}_r^{(i)} = \bigl[s_{r;u,v}^{(i)}\bigr]\;, 0 \le u \le n_s-1,\; 0 \le v \le n_t-1$, 
where \(u\) indexes the frequency bin and \(v\) indexes the time slice.


Depending on the available computational resources, each spectrogram is further downsampled along the time axis to obtain $\widetilde{\mathbf{S}}_r^{(i)} \in \mathbb{R}^{n_s \times n_t'}$, where $n_t'$ is the length of the downsampled time axis. This sequence is partitioned into an input segment of length \(n_{t,\mathrm{in}}\) and an output segment of length \(n_{t,\mathrm{out}}\), with $n_t' = n_{t,\mathrm{in}} + n_{t,\mathrm{out}}$.


The \textbf{spectrum prediction problem} can therefore be formulated as follows. Given a record characterized by site $ST$, center frequency $f_i$, timestamp $T_r^{(i)}$, receiver gain $G$, and frequency-bin index $u$, the objective is to predict the next \(n_{t,\mathrm{out}}\) PSD values from the first \(n_{t,\mathrm{in}}\) observed values. Equivalently, the prediction task may be written as
\[
\underbrace{\bigl(\widetilde{s}_{r;u,v}^{(i)}\bigr)_{v=0}^{n_{t,\mathrm{in}}-1}}_{\text{input}}
\;\longmapsto\;
\underbrace{\bigl(\widetilde{s}_{r;u,v}^{(i)}\bigr)_{v=n_{t,\mathrm{in}}}^{n_t'-1}}_{\text{target}}.
\]


\vspace{-0.05in}
\section{Methodology and Architecture}

\label{sec:methodology_and_architecture_lsm}

In this section, we outline the methodology for data preprocessing and model development aimed at predicting spectrum usage. It begins with a detailed description of the data preprocessing steps, followed by an in-depth discussion of the development process of the spectrogram prediction model, leveraging LLM architectures.

\vspace{-0.05in}
\subsection{Data Preprocessing}
\label{subsec:data_preprocessing_lsm}

The data preprocessing step serves two key purposes: (1) it enables the data to be efficiently handled within the constraints of available computational resources, and (2) it ensures compatibility with modern machine learning and deep learning libraries~\cite{tensorflow15, Paszke2019pytorch}. 

The bandwidth and duration of each sample collection mentioned in Section~\ref{subsec:prob_des_lsm} are fixed for all samples within this spectrum dataset, specifically: $B=20\: \mathrm{MHz}$ and $d=1s$. Consequently, each spectrum sample is represented by an IQ vector of size $N = 20$,$000$,$000$. Utilizing STFT, this raw spectrum sample in the  time-domain can be converted into a time-frequency matrix. We employ a non-overlapping STFT with a frequency bin size of $n_s=256$. As a result, the dimension of the resulting time-frequency matrix is $256 \times 78$,$125$. Finally, applying the magnitude-to-decibel-milliwatt ($dBm$) conversion to each element of the time-frequency matrix yields a two-dimensional spectrogram, where the number of time slices, $t$, equals $78$,$125$. Due to computational limitations in handling such long sequences with machine learning models, particularly during GPU-based training, further downsampling of these spectrograms is required. To address this, we employ a two-step downsampling procedure.

First, each frequency bin of the spectrogram is downsampled using $1D$ max-pooling. Given that $t=78$,$125$ and $d=1s$, each element of a frequency bin sequence corresponds to $12.8\mu s$. We select a window size and stride of $25$ for the max-pooling operation, resulting in the selection of the maximum PSD value within every $320\mu s$. Thus, within every $320\mu s$ interval, the maximum PSD value, which typically represents the most significant metric for spectrum analysis, is retained for further analysis. Upon completion of this operation, the spectrogram dimensions are reduced to $256 \times 3$,$125$.

Second, we employ the trimmed mean method to further reduce the sequence length obtained from the max-pooling step. Trimmed mean is a robust statistical technique, widely recognized in the literature for its effectiveness in reducing the influence of outliers, particularly in time-series analyses~\cite{oliveira2023trimmed, kocak4937461trimmed}. By applying this method, the spectrogram dimensions are effectively reduced from $256 \times 3$,$125$ to $256 \times 256$, greatly facilitating subsequent analysis and machine learning tasks. Specifically, around $12$ consecutive values in each spectrogram frequency bin sequence are used to calculate each trimmed mean value. In these experiments, the first $n_{tin}=128$ elements of each sequence are used as model input, while the remaining $n_{tout}=128$ elements are predicted by the model, resulting in a total sequence length of $n'_t = 256$.

As a result, each PSD value in the spectrogram now represents a time interval of approximately $3.91\mathrm{ms}$. On the other hand, in the frequency domain, the total $20\:\mathrm{MHz}$ bandwidth is divided into $256$ equal frequency bins. This implies that each frequency bin occupies a bandwidth of $78.125\:\mathrm{kHz}$. Therefore, every PSD value in the dataset corresponds to a time-frequency resolution of $3.91\mathrm{ms} \times 78.125\: \mathrm{kHz}$.

Blending both max-pooling and trimmed mean offers twofold benefits: First, it ensures that outliers resulting from power leakage across adjacent frequency bins are effectively excluded from the final dataset, a limitation that could arise if only max-pooling were used. Second, it preserves all genuine high-power activities, which are of primary interest in spectrum sensing applications.

\begin{table}[t]
\centering
\caption{Tokenization Scheme for Dataset Parameters}
\label{tab:tokenization_scheme}
\resizebox{0.98\linewidth}{!}{%
\begin{tabular}{|l|c|c|c|}
\hline
\textbf{Parameter} & \textbf{Original Range} & \textbf{Token Range} & \textbf{Begin/End Tags} \\
\hline
PSD (dBm)         & $-118$ to $-18$        & $1$ to $101$      & $126$, $127$ \\
Gain (dB)         & $0$ to $79$            & $1$ to $80$       & $118$, $119$ \\
Frequency (MHz)   & $54$ to $990$          & $1$ to $33$       & $114$, $115$ \\
frequency bin        & $0$ to $255$           & $[0,15],[0,15]$   & $116$, $117$ \\
Day of Week       & Sat to Fri             & $1$ to $7$        & $122$, $123$ \\
Day               & $1$ to $31$            & $1$ to $31$       & $102$, $103$ \\
Month             & Jan to Dec             & $1$ to $12$       & $104$, $105$ \\
Year              & $2016$ to $2035$       & $1$ to $20$       & $106$, $107$ \\
Hour              & $0$ to $23$            & $1$ to $24$       & $108$, $109$ \\
Minute            & $0$ to $59$            & $1$ to $60$       & $110$, $111$ \\
Second            & $0$ to $59$            & $1$ to $60$       & $112$, $113$ \\
\hline
\end{tabular}
}
\vspace{-0.25in}
\end{table}

\vspace{-0.05in}
\subsection{RF Tokenization}
\label{subsec:data_tokenization_lsm}

Tokenization plays a critical role in enabling compatibility with conventional language models. In this section, we outline the detailed methodology used to tokenize the spectrum sensing dataset.

So far, we have converted the raw spectrum data into a $2D$ spectrogram and downsampled it to the shape of $n_s \times n_t^{\prime}$, where $n_s = n_t^{\prime} = 256$. Each spectrogram contains floating-point values that represent the PSD for a specific frequency bin and time span in $dBm$. For tokenization, the first step is to discard the fractional parts and convert the PSD values to integers. This simplification facilitates the tokenization process while having minimal impact on the model's ability to learn and predict spectrum patterns effectively.

\begin{figure*}[t!]
	\centering
	\subfloat[LSM-Gemma]{
		\includegraphics[height=5in]{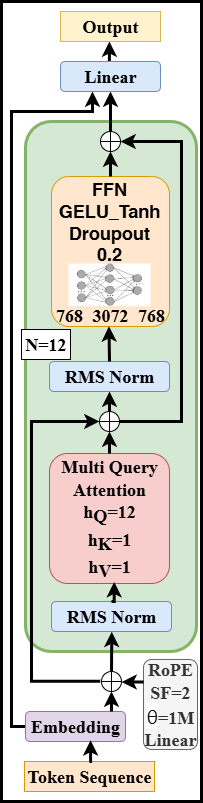}\label{fig:gemma}
	}
	\subfloat[LSM-GPT]{
		\includegraphics[height=5in]{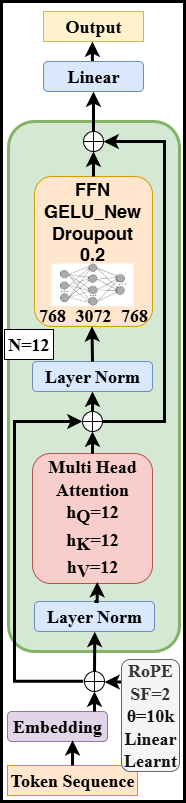}\label{fig:gpt}
	}
	\subfloat[LSM-LLaMA]{
		\includegraphics[height=5in]{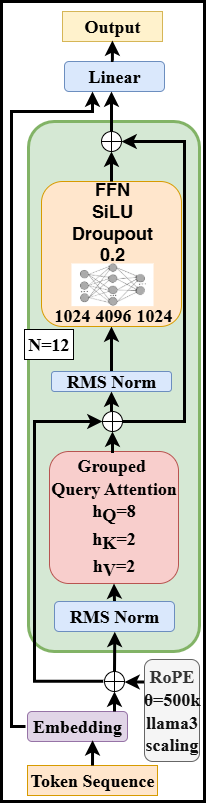}\label{fig:llama}
	}
    \subfloat[LSM-Mistral]{
		\includegraphics[height=5in]{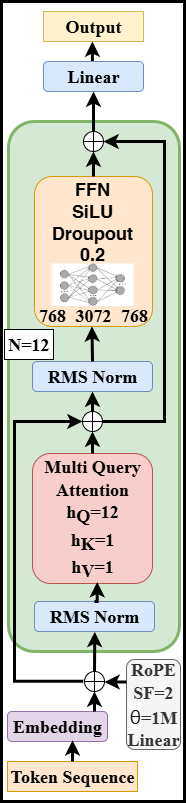}\label{fig:mistral}
	}
    \subfloat[LSM-Phi]{
		\includegraphics[height=5in]{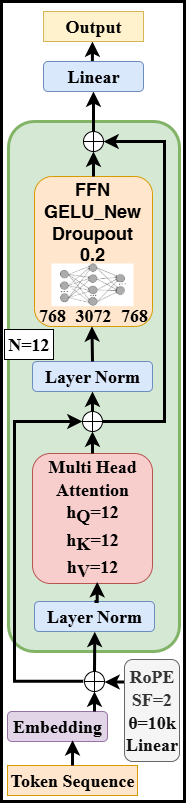}\label{fig:phi}
	}    

	\vspace{-0.05in}
	\caption{\it LSM Architectures.}\label{fig:model_arch}
	\vspace{-0.3in}
\end{figure*}

After omitting the fractional parts, the PSD values across the entire dataset range from $-118\mathrm{dBm}$ to $-18\mathrm{dBm}$. Each of these $101$ distinct PSD levels is mapped to a unique token ranging from $1$ to $101$. This mapping provides a $1\mathrm{dB}$ resolution, ensuring that each PSD value in the dataset corresponds to a dedicated token. As a result, the spectrogram sequences can be directly converted into sequences of discrete tokens suitable for language-model-based processing.

Similarly, we tokenize four other critical parameters of the dataset: gain, frequency, frequency bin, and timestamp. The receiver front-end of the USRP N310 supports gain values ranging from $0$ to $79\mathrm{dB}$. To represent these, we assign tokens from $1$ to $80$. For frequency, the dataset includes $33$ distinct center frequencies, each assigned a unique token from $1$ to $33$.

For the frequency bins, which contains $256$ possible values, we adopt a two-token representation instead of assigning a single token to each frequency bin. Inspired by the base-16 (hexadecimal) numbering system, each frequency bin index is represented using two tokens, each ranging from $1$ to $16$. This representation significantly reduces the overhead of the token space, preventing unnecessary expansion that would result from assigning a unique token to each frequency bin instance.

Since the timestamp is represented as a $64$-bit integer, it is not feasible to encode it directly as a single token within any practical token space. Instead of using the raw timestamp, we extract several interpretable components for use in the dataset: day of the week, date, month, year, hour, minute, and second. Each of these components is then numerically tokenized based on the range of its possible values.

The tokenized dataset consists of two primary components: (1) spectrum sequence data, which includes the PSD vectors in $dBm$, and (2) metadata, which comprises gain, frequency, frequency bin, and timestamp information. To enable integration of the core dataset with its associated metadata, we introduce a set of special tags. For this purpose, we reserve token values from $102$ to $127$ to define these tags. These reserved tokens include the necessary beginning and end tags for each of the parameters discussed above, allowing for structured and unambiguous parsing of the combined token sequence. Reserving dedicated tags separate from the core dataset ensures \textbf{clear delineation between different components of the input}, facilitating easier separability for the transformer's attention scheme.

In Table~\ref{tab:tokenization_scheme} a concise overview of the complete tokenization scheme is provided. For each parameter, the original value range, corresponding tokenized range, and the associated beginning and end tags are listed. Additionally, token $0$ is reserved for data padding, while tokens $124$ and $125$ are specifically reserved to indicate the beginning and end of the entire metadata block.

The overall data formatting becomes simple and structured with this tokenization methodology. All relevant metadata can be placed at the beginning, followed by the tokenized spectrum sequence corresponding to that metadata. This approach encapsulates the entire dataset into a unified structure, making it readily compatible with standard LLM architectures.

Therefore, the entire vocabulary space ranges from $0$ to $127$, meaning that the dataset can be fully represented using a single-byte signed integer format. This brings two key benefits: (1) the model's output layer only needs to project onto $128$ output neurons, which \textbf{significantly reduces the model size and improves memory efficiency on the GPU}; and (2) the core dataset can be transferred between high-performance computing (HPC) platforms for training and evaluation much faster than traditional large-vocabulary datasets due to its compact representation. 

After the formation of tokenized metadata, each sequence is extended with additional $36$ values that serve as a metadata header, resulting in a total sequence length of $292$. The complete dataset comprises \textbf{approximately $8.43$ billion tokens}. For fine-tuning, we additionally use a separate smaller dataset collected from the auxiliary site, which contains a total of $246$ million tokens.

\vspace{-0.05in}
\subsection{LSM Architectures}
\label{subsec:model_architecture_lsm}
We select five well-known open-source LLMs for training and evaluating the proposed spectrum sensing dataset. Several key criteria guided this selection. First, we prioritize decoder-only transformer models that are fully open-source to ensure complete control over their implementation and customization. Second, we aim to cover a diverse range of transformer architectural characteristics across the selected models. Third, we consider computational feasibility to ensure that the models could be efficiently trained and evaluated using the available computing resources we have for this work.

Based on these criteria, we select the following five models: (1) Gemma-2B from Google DeepMind~\cite{Gemma24}, (2) GPT-2 from OpenAI~\cite{GPT19}, (3) LLaMA-7B from Meta~\cite{LLaMA23, huggingface2025llama}, (4) Mistral-7B from Mistral AI~\cite{Mistral23}, and (5) Phi-1 from Microsoft~\cite{Phi23} All five models are available through the Hugging Face \texttt{transformers} library~\cite{wolf2019huggingface}. 



In Figs.~\ref{fig:model_arch}, we present the models utilized for training and evaluating the spectrum dataset. Each model represents a scaled implementation of its original architecture, modified as necessary to accommodate task and computational constraints. However, the key properties of the original models are preserved. Consequently, we prepend each model name with the superscript ``LSM" to denote its adaptation for large-scale spectrum modeling.

To provide a detailed explanation of the architectural development, we begin by considering the architecture of a conventional LLM, namely GPT-2~\cite{GPT19}. OpenAI developed GPT-2 with $48$ transformer layers, each incorporating multi-head attention (MHA) with $25$ heads, where the query (Q), key (K), and value (V) representations are $1600$-dimensional and partitioned into $64$-dimensional subspaces per head. The outputs of these $48$ MHA layers are then fed into a feed-forward network (FFN) with an input dimension of $1600$. The hidden layer of this FFN has a dimensionality of $6400$, and the output layer returns to $1600$ dimensions This FFN is ultimately followed by a linear neural network that produces the probability distribution over tokens for a given input sequence.

\begin{figure}[t!]
	\centering
		\includegraphics[width=0.49\textwidth]{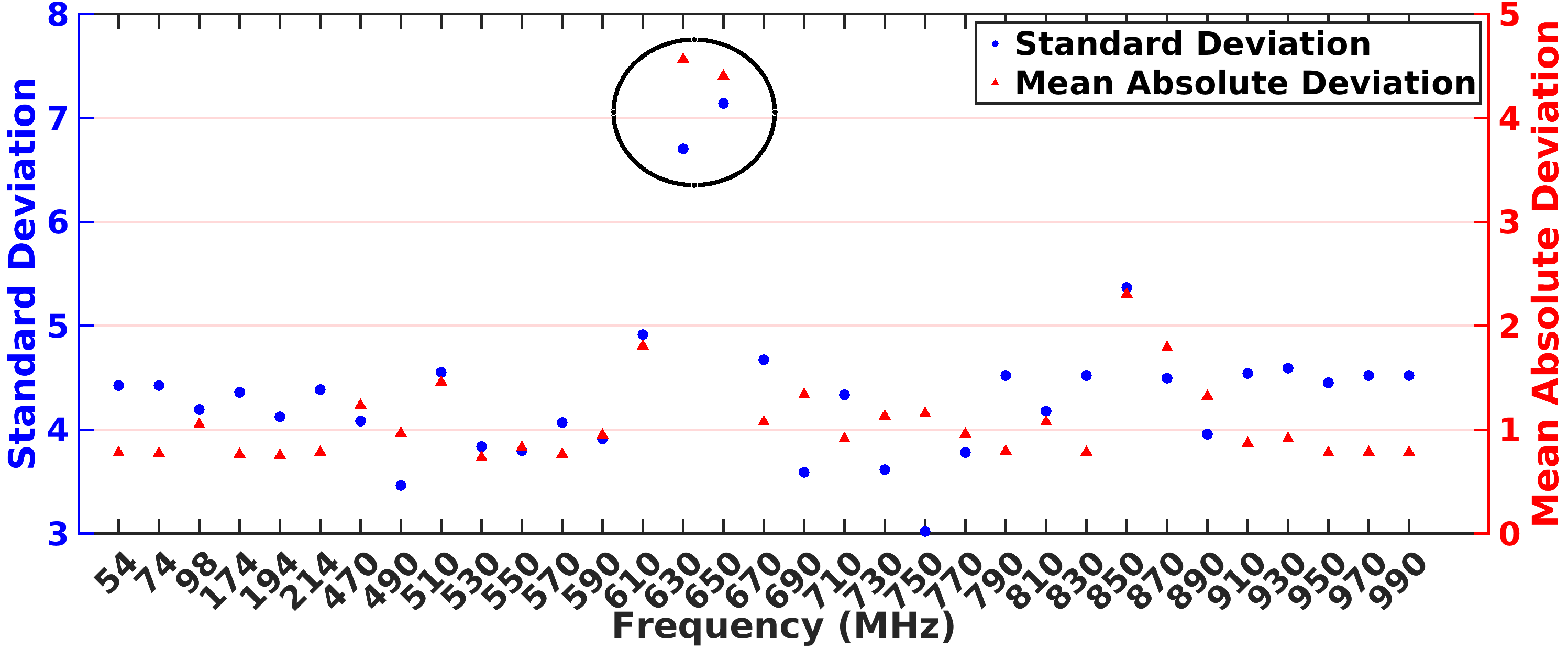}
		\vspace{-0.2in}
	\caption{\it Average SD \& average MAD across the entire dataset.}\label{fig:std_mad}
	\vspace{-0.2in}
\end{figure}

\begin{figure*}[t!]
	\centering
		\includegraphics[width=0.95\textwidth]{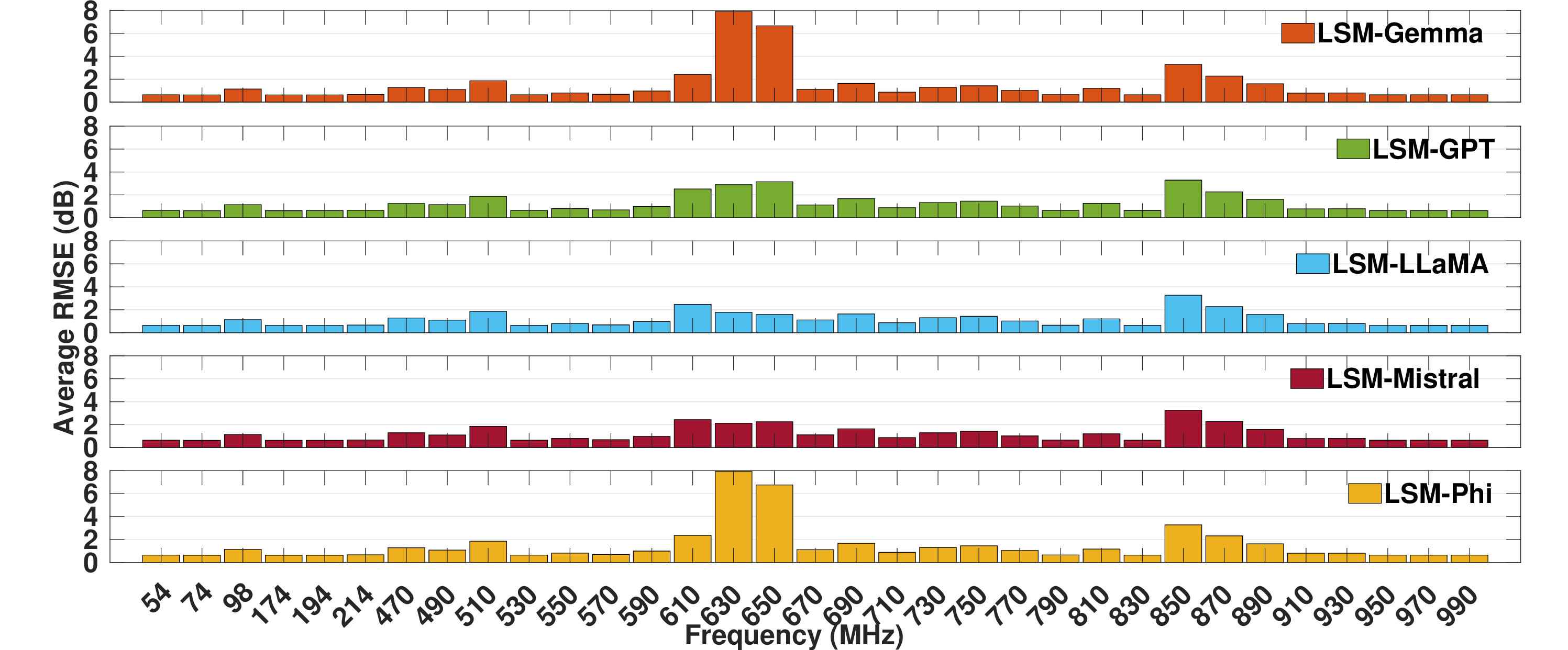}
		\vspace{-0.05in}
	\caption{\it Average RMSE for the five LSM models.}\label{fig:rmse_subplot}
	\vspace{-0.2in}
\end{figure*}

In this work, we adopt a similar yet downscaled architecture for the LSM variant of GPT. Specifically, LSM-GPT (Fig.~\ref{fig:gpt}) incorporates $12$ MHA layers, each with $12$-dimensional Q, K, and V tensors. The FFN in LSM-GPT has an input dimension of $768$, a hidden layer with $3072$ neurons, and an output dimension of $768$. The normalization and activation layers are retained as in the original GPT-2 architecture. All components of LSM-GPT are derived from the GPT-2 small configuration.

This architectural design is chosen to ensure that the model can be effectively loaded and trained using available computational resources. All training and validation tasks are performed on a single NVIDIA H100 GPU with $80$GB of GPU memory. The designed architecture and dataset are well-suited to fit within these hardware constraints.

Similarly, for developing LSM-Gemma, we take Gemma-2B~\cite{Gemma24} as a reference and align its architecture with LSM-GPT as closely as possible. Gemma-2B consists of $18$ multi-query attention (MQA) layers, each comprising $8$ query (Q), $1$ key (K) and $1$ value (V) tensors. Its feed-forward network (FFN) contains $2,048$, $32,768$, and $2,048$ neurons in the input, hidden, and output layers, respectively.

Building on this architecture and maintaining consistency with LSM-GPT, we design the LSM-Gemma (Fig.~\ref{fig:gemma}) with $12$ MQA layers, each containing $12$ Q, $1$ K and $1$ V tensors. The FFN in LSM-Gemma follows a $768 \rightarrow 3072 \rightarrow 768$ structure corresponding to the input, hidden, and output neurons, respectively. 

In an analogous manner, LSM-LLaMA (Fig.~\ref{fig:llama}), LSM-Mistral (Fig.~\ref{fig:mistral}), and LSM-Phi (Fig.~\ref{fig:phi}) are developed, adhering to the same design principles to ensure architectural consistency across the LSM variants. Among these, LSM-LLaMA differs in its attention mechanism. It adopts grouped-query attention (GQA), using a $2$-group configuration. In this setup, a total of $8$ query heads are divided into two groups, each group sharing a single key ($K$) and a single value ($V$) tensor. As a result, the model employs two $K$ tensors and two $V$ tensors in total.

In contrast, LSM-Mistral leverages the MQA mechanism, similar to LSM-Gemma. Finally, LSM-Phi employs multi-head attention (MHA), analogous to LSM-GPT.

We achieve two key benefits by maintaining architectural similarities across all LSM models as much as possible. First, it maintains architectural similarity across all models, except for the core attention mechanisms inherited from their respective base models. This allows performance comparisons to be made independently of other factors. Second, this approach ensures that the computational resources available for this work are properly and consistently utilized for each model.

Among these models, LSM-Gemma, LSM-LLaMA, and LSM-Mistral employ RMS normalization, while LSM-GPT and LSM-Phi utilize layer normalization. For activation functions, we observe three types across the models: GELU-Tanh, GELU-New, and SiLU. LSM-Gemma uses GELU-Tanh, whereas LSM-GPT and LSM-Phi adopt GELU-New. LSM-LLaMA and LSM-Mistral use SiLU activation.

Both normalization and activation settings are kept exactly the same as in their respective base models. Furthermore, to mitigate overfitting, we apply a $20\%$ dropout rate to both the feed-forward network and the attention layers across all models.

In addition, rotary positional encoding (RoPE) is implemented across all five models. RoPE helps the models learn positional dependencies in long-sequence datasets. We retain these settings as defined in their respective base models to ensure consistency. An important point to note is that LSM-GPT and LSM-Phi are almost identical except for their RoPE configurations. LSM-GPT adopts a learned RoPE scheme with linear position placement, whereas LSM-Phi employs the standard linear RoPE scheme.

\vspace{-0.05in}
\subsection{Model Training and Data Organization Strategy}
\label{subsec:training_and_org_lsm}

Before discussing the details of model training, we first examine the dataset used in this study. To understand its distribution, we analyze two statistical measures: the standard deviation (SD) and mean absolute deviation (MAD). In Fig.~\ref{fig:std_mad}, we present these statistics for each frequency band in the dataset.

As shown in the figure, the data corresponding to $630\:\mathrm{MHz}$ and $650\:\mathrm{MHz}$ exhibit significantly higher SD and MAD values (highlighted with a black circle) compared to the other $31$ bands. Based on this observation, we divide the dataset into two major categories: highly dispersed (HD) data and regular data. Specifically, the entire dataset corresponding to $630\:\mathrm{MHz}$ and $650\:\mathrm{MHz}$ is classified as HD data, while data from the remaining $31$ frequency bands are categorized as regular data.

Each frequency band contains approximately an equal number of samples. Consequently, the HD data represent roughly $6\%$ of the total dataset. Due to the high SD and MAD in this portion, it requires special consideration during training. To improve the learning effectiveness for HD data, we conduct a series of experiments with different loss functions.

Instead of using the default cross-entropy (CE) loss, we employ a hybrid loss function that combines CE and RMS loss, defined as follows:
\vspace{-0.05in}
\begin{equation}
\vspace{-0.05in}
    L_{total} = \alpha L_{CE} + (1-\alpha) L_{RMS},
    \label{eqn:loss}
\end{equation}
where $\alpha$ is the mixing coefficient that balances the contribution of CE loss and RMS loss. We determine $\alpha = 0.1$ to be effective for the HD dataset. For the regular dataset, we retain the default CE loss only, that is, $\alpha = 1$, since it yields sufficiently accurate results in all models.

\begin{table}[t]
\centering
\caption{Training Resource Summary of LSM Models}
\vspace{-0.05in}
\label{tab:lsm_training_resources}
\begin{tabular}{|l|r|r|}
\hline
\textbf{Model} & \textbf{Total Parameters} & \textbf{H100 GPU Hours} \\
\hline
LSM-Gemma   & 292,787,712 & 76.66 \\
LSM-GPT     & 170,757,120 & 82.16  \\
LSM-LLaMA   & 402,966,528 & 74.93 \\
LSM-Mistral & 200,971,776 & 67.04 \\
LSM-Phi     & 170,468,608 & 90.49  \\
\hline
\end{tabular}
\vspace{-0.25in}
\end{table}

In Table~\ref{tab:lsm_training_resources}, we present the total number of parameters and the overall training time on an $H100$ GPU for each model. For each case, we develop two separate models: one trained on the HD dataset and the other on the regular dataset. The parameter counts and training times reported in the table represent the combined values of both models for each LSM variant. Although LSM-GPT and LSM-Phi employ MHA architecture, they have lower numbers of parameters ($170M$ each) than the other models. This is due to the design of their feedforward layers and attention projections. Specifically, the MHA variants use more compact feedforward widths and do not require the expanded key--value projection structures introduced in GQA and MQA models. As a result, despite the higher computational overhead of MHA during training, LSM-GPT and LSM-Phi maintain the smallest parameter counts among the five LSMs. LSM-LLaMA has the highest parameter count ($402M$) due to the higher FFN size than the others.

Finally, to prevent temporal leakage in the time-series data, the dataset is split chronologically rather than randomly. Specifically, the data is organized such that $\text{time}_{\text{train}} < \text{time}_{\text{val}} < \text{time}_{\text{test}}$, ensuring that validation and test samples occur strictly after the training samples.

\vspace{-0.05in}
\section{Performance Evaluations}
\label{sec:evaluation}

In this section, we present the evaluation results of the LSM models on the spectrum sensing dataset.

\vspace{-0.1in}
\subsection{Benchmarking Against Time-Series Models}

In Table~\ref{tab:lstm}, the benchmarking results are summarized for recurrent and time-series transformer models on the HD dataset ($630$ MHz and $650$ MHz bands), including model size and training cost on a single NVIDIA H100 GPU.

We first evaluate stacked LSTM architectures by varying the number of layers ($N_L$) and the size of the hidden state ($h_S$). Increasing the hidden size from $256$ to $512$ improves RMSE while increasing parameters from $1.1$M to $3.8$M. Further scaling to $h_S=3072$ expands the model to $123$M parameters and increases training time to $34.93$ GPU hours without improving prediction accuracy. Larger hidden sizes ($h_S \ge 4096$) or deeper networks lead to unstable training where the model converges to nearly constant outputs due to gate saturation and vanishing gradients. As a result, the RMSE values for these configurations have been omitted.

Increasing network depth shows a similar trend: raising depth from $N_L=2$ to $N_L=12$ at $h_S=512$ increases parameters from $3.8$M to $24.8$M and training time from $8.95$ to $47.23$ GPU hours while again converging to constant-output behavior.

For comparison, we also evaluate a transformer-based Time-GPT model~\cite{Garza2023} with $12$ layers and $12$ attention heads ($198$M parameters). This model requires $61.68$ GPU hours of training and achieves RMSE values of $8.8154$ and $10.2235$ for the $630$ MHz and $650$ MHz bands, respectively. Based on the overall analysis, the configuration with $N_L=2$ and $h_S=512$ is selected as the non-LSM baseline.

\begin{table}[t]
\centering
\caption{LSTM (top rows) and Time-GPT (bottom row) model configurations, compute cost, and performance}
\vspace{-0.05in}
\label{tab:lstm}
\begin{tabular}{|c|c|c|c|c|c|}
\hline
\multirow{2}{*}{\textbf{$N_L$}} 
& \multirow{2}{*}{\textbf{$h_S$/$N_H$}} 
& \multirow{2}{*}{\textbf{Parameters}} 
& \multirow{2}{*}{\textbf{H100 (hrs)}} 
& \multicolumn{2}{c|}{\textbf{RMSE}} \\ \cline{5-6}
& & & & \textbf{630 MHz} & \textbf{650 MHz} \\ \hline
$2$  & $256$  & 1,118,336     & 6.98  & 9.1624 & 10.1680 \\ 
$2$  & $512$  & 3,776,640     & 8.95  & 7.2793 & 6.6877 \\ 
$2$  & $1024$ & 15,974,528    & 10.38 & 7.7372 & 6.7836 \\ 
$2$  & $3072$ & 123,224,192   & 34.93 & 8.9754 & 9.6315 \\ 
$2$  & $4096$ & 214,597,760   & 46.83    & -- & -- \\ 
$4$  & $512$  & 7,979,136     & 13.42 & -- & -- \\ 
$8$  & $512$  & 16,384,128    & 26.43 & -- & -- \\ 
$12$ & $512$  & 24,789,120    & 47.23 & -- & -- \\ \hline
$12$  & $12$  & 198,327,552     & 61.68  & 8.8154 & 10.2235 \\ \hline
\end{tabular}
\vspace{-0.20in}
\end{table}

\vspace{-0.1in}
\subsection{LSM RMSEs and MAEs}

To evaluate the performance of different transformer-based architectures in spectrum prediction, we compute the RMSE in dB across all $33$ center frequencies in the sub-GHz band. In Fig.~\ref{fig:rmse_subplot}, we present a comparative analysis of RMSE values for the five models.

The results show that LSM-LLaMA and LSM-Mistral demonstrate the best consistency, with standard deviations of RMSE equal to $0.63$ and $0.66$, respectively. In comparison, LSM-Gemma, LSM-GPT, and LSM-Phi exhibit higher variability, with standard deviations of $1.62$, $0.77$, and $1.63$, respectively. Furthermore, LSM-Mistral achieves the best overall performance by maintaining the lowest maximum RMSE across bands, measured at $3.25$ dB. LSM-LLaMA follows closely, keeping all RMSE values less than or equal to $3.26$ dB.

LSM-Gemma and LSM-Phi struggle the most at the $630\:\mathrm{MHz}$ and $650\:\mathrm{MHz}$ bands, which we classify as part of the HD dataset. We observe RMSE values of $7.90$ dB and $7.93$ dB at $630\:\mathrm{MHz}$ for LSM-Gemma and LSM-Phi, respectively. At $650\:\mathrm{MHz}$, their RMSE values are $6.65$ dB and $6.76$ dB, respectively. These frequency bands are used for wireless broadband services, and the dynamic nature of cellular user activity significantly influences spectrum usage in these bands.

In contrast, LSM-LLaMA and LSM-Mistral demonstrate strong robustness in learning the PSD characteristics of these bands. LSM-LLaMA maintains RMSE values below $1.8$ dB, while LSM-Mistral achieves RMSE values below $2.25$ dB. As a baseline comparison, the best-performing time-series model achieves RMSE values of $7.28$ dB at $630$ MHz and $6.69$ dB at $650$ MHz.
These errors are substantially higher than those obtained by the best-performing LSMs, namely LSM-LLaMA, LSM-Mistral, and LSM-GPT, on the HD dataset, highlighting the advantage of transformer-based architectures for modeling highly dynamic spectrum activity. Although the computational overhead of the LSTM, including model parameters and H100 GPU training hours, is up-to $170$ times lower than that of LSMs, the resulting prediction accuracy renders it less effective for high-fidelity spectrum modeling in this setting.

Interestingly, LSM-LLaMA and LSM-Mistral, which use GQA and MQA, consistently outperform the MHA-based LSM-GPT and LSM-Phi. Although MHA is theoretically more expressive, its independent key--value projections add redundancy and computational overhead, leading to overfitting on large spectrum datasets. In contrast, GQA and MQA share key value representations, improving efficiency, reducing variance, and enhancing the capture of dominant PSD dynamics, particularly in highly dynamic licensed bands. Consequently, LSM-LLaMA and LSM-Mistral achieve superior robustness and generalization.

\begin{figure}[t!]
	\centering
		\includegraphics[width=0.49\textwidth]{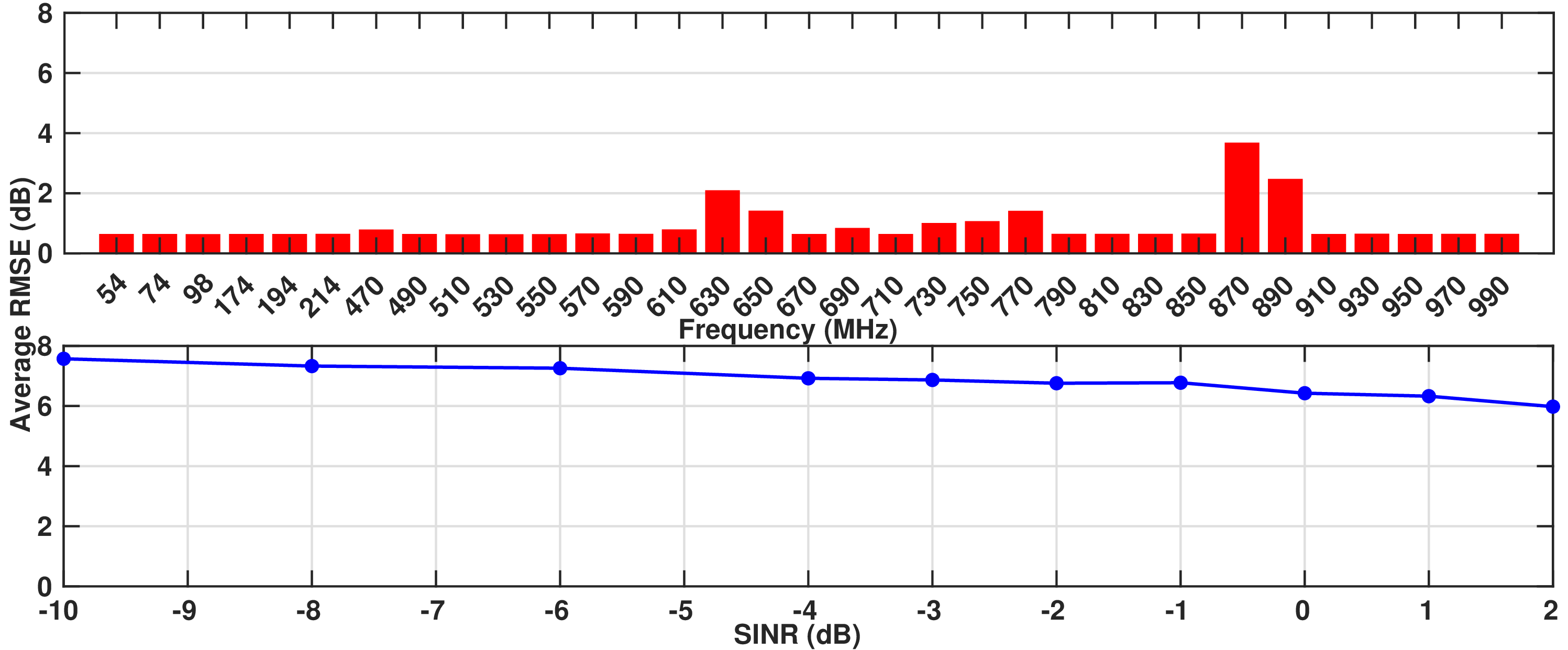}
		\vspace{-0.2in}
	\caption{Average RMSE on the auxiliary site dataset (top) and simulation dataset with varying SINR (bottom) for the fine-tuned model.}\label{fig:finetuning-combined}
	\vspace{-0.05in}
\end{figure}

\begin{table}[t]
\centering
\caption{Weighted Kappa measures of LSM models}
\label{tab:weighted_kappa}
\begin{tabular}{|l|r|r|}
\hline
\textbf{Model} & \textbf{Overall} & \textbf{HD Dataset} \\
\hline
LSM-Gemma   & 0.92 & 0.73 \\
LSM-GPT     & 0.93 & 0.91  \\
LSM-LLaMA   & 0.94 & 0.95 \\
LSM-Mistral & 0.94 & 0.93 \\
LSM-Phi     & 0.92 & 0.73  \\
\hline
\end{tabular}
\vspace{-0.25in}
\end{table}

We also measure the MAE for each predicted PSD sequence and observe trends consistent with those of RMSE. Except for the $630\:\mathrm{MHz}$ and $650\:\mathrm{MHz}$ bands, all models maintain more than $89\%$ of sequence predictions with MAE below $5$ dB. LSM-LLaMA maintains this level for more than $97\%$ of the sequences, while LSM-Mistral achieves more than $94\%$ of the predictions with MAE below $5$ dB. However, LSM-Gemma and LSM-Phi achieve this threshold for only $69\%$ of the test sequences. 

From a system perspective, lower RMSE and MAE corresponds to more accurate estimation of future spectrum power levels. Such improvements can enable spectrum management systems to better anticipate channel occupancy and interference conditions, which is essential for proactive channel selection and spectrum allocation in DSA frameworks. Consequently, improved prediction accuracy can translate into more reliable spectrum access decisions in practical deployments.

\vspace{-0.05in}
\subsection{Weighted Kappa Measures}

Since the PSD values of this dataset range from $-118$ dBm to $-18$ dBm, the entire dataset can be represented as $101$ discrete classes. The performance of the model can therefore be evaluated as a multi-class classification problem. We employ Cohen's weighted kappa~\cite{Kohen1960coefficient} to assess multi-classification performance. In this approach, the weights of disagreement are determined on the basis of the deviation between the predicted and target PSD values. The deviation is measured as the absolute difference in dB.

In Table~\ref{tab:weighted_kappa}, we report the weighted kappa values for the five models. Across the overall dataset, all models achieve values greater than $0.91$, demonstrating their strong ability to predict PSDs throughout the sub-GHz band. For the HD dataset ($630\:\mathrm{MHz}$ and $650\:\mathrm{MHz}$ bands), the measure drops to $0.73$ for the LSM-Gemma and LSM-Phi models, which is consistent with their RMSE performance. In contrast, all other models achieve values above $0.90$ in the HD dataset, highlighting their robustness in these challenging bands.

\vspace{-0.05in}
\subsection{Fine-Tuning the Pretrained Model}

Large datasets may not be available for all sites of interest. To assess the generalization capabilities of LSMs, we evaluate their fine-tuning performance on two separate datasets. The first dataset is collected from the auxiliary site described in Section~\ref{subsec:datacol_lsm}. To ensure it is sufficiently smaller than the core dataset, we use a subset containing $246$ million tokens, which corresponds to $2.92\%$ of the core dataset in terms of token count.

The second dataset is a simulation data generated using MATLAB's $5G$ Toolbox for downlink transmission at $630\:\mathrm{MHz}$. The simulated signal is passed through an additive white Gaussian noise (AWGN) channel with varying signal-to-interference-plus-noise ratios (SINRs). Both fine-tuning experiments are conducted using the LSM-Mistral pretrained model, which has been identified as one of the most successful LSM variants.

In Fig.~\ref{fig:finetuning-combined}, we present the results for both fine-tuned models. The top plot shows the RMSE for the auxiliary site dataset, where the average RMSE across all bands ranges from $0.63$ dB to $3.68$ dB. Compared to the performance of LSM-Mistral on the core dataset in Fig.~\ref{fig:rmse_subplot}, we observe less than a $1$ dB increase in RMSE across all bands except at $870\:\mathrm{MHz}$, where the average RMSE increase is $1.4$ dB (final value is $3.68$ dB). Since the model maintains an average RMSE below $3.70$ dB for all bands, even with such a small dataset, the fine-tuned LSM-Mistral achieves accuracy comparable to that of the core model.

In the bottom subplot of Fig.~\ref{fig:finetuning-combined}, we show the average RMSE values with respect to varying SINR levels. The SINR ranges from $-10$ dB to $2$ dB. At $-10$ dB, the RMSE is $7.57$ dB, and as the SINR increases, the RMSE gradually improves, reaching $5.97$ dB at $2$ dB SINR. Despite being evaluated on a completely different dataset, LSM maintains an RMSE below $7.6$ dB across all SINR values, demonstrating the robustness of the model on diverse datasets.



\vspace{-0.05in}
\section{Conclusion}
\label{sec:lsm_conclusion}

In this paper, we introduce foundational large spectrum models (LSMs), novel deep learning frameworks for wideband power spectral density prediction using large-scale transformer-based models. The results demonstrate that the decoder-only transformer-based LSM architectures can achieve acceptable accuracy in spectrum forecasting, relative to recurrent baselines. Furthermore, site-specific and application-specific datasets can greatly benefit from the pretrained models for fine-tuning. To the best of our knowledge, this work is the first to systematically apply general-purpose LLM architectures to real-world wideband RF data at scale, demonstrating their effectiveness in the spectrum domain without the need for extensive task-specific architectural modifications. The proposed methodology establishes a foundation for future research on applying LSMs to wireless signal intelligence, particularly in dynamic spectrum access and spectrum management for 6G and beyond wireless systems. 


\vspace{-0.1in}
\section*{Acknowledgments}

This work is supported in part by NSF grants CNS-2212050, CNS-2030272, and OAC-2321699 and utilized the Nebraska Research Data Storage (NRDStor), operated by the Holland Computing Center (HCC) at the University of Nebraska--Lincoln and supported by NSF grant OAC-2232851. We thank Garhan Attebury for his support in NEXTT and Michael Fay and Prashant Subedi for their assistance with data collection.

\bibliographystyle{IEEEtran}
\bibliography{reference_lsm_bib_short, vuran-cpn-works} 

\begin{thebibliography}{10}
\providecommand{\url}[1]{#1}
\csname url@samestyle\endcsname
\providecommand{\newblock}{\relax}
\providecommand{\bibinfo}[2]{#2}
\providecommand{\BIBentrySTDinterwordspacing}{\spaceskip=0pt\relax}
\providecommand{\BIBentryALTinterwordstretchfactor}{4}
\providecommand{\BIBentryALTinterwordspacing}{\spaceskip=\fontdimen2\font plus
\BIBentryALTinterwordstretchfactor\fontdimen3\font minus \fontdimen4\font\relax}
\providecommand{\BIBforeignlanguage}[2]{{%
\expandafter\ifx\csname l@#1\endcsname\relax
\typeout{** WARNING: IEEEtran.bst: No hyphenation pattern has been}%
\typeout{** loaded for the language `#1'. Using the pattern for}%
\typeout{** the default language instead.}%
\else
\language=\csname l@#1\endcsname
\fi
#2}}
\providecommand{\BIBdecl}{\relax}
\BIBdecl

\bibitem{IoTA24}
S.~Sinha, ``{State of IoT 2024: Num. of connected IoT devices growing 13\% to 18.8B globally},'' \url{https://bit.ly/4l36mLs}, 2024, accessed: 2025-06-21.

\bibitem{Wang6G}
C.-X. Wang \emph{et~al.}, ``On the road to {6G}: Visions, requirements, key technologies, and testbeds,'' \emph{IEEE Communications Surveys and Tutorials}, vol.~25, no.~2, pp. 905--974, 2023.

\bibitem{Akyildiz2006NeXt}
I.~F. Akyildiz, W.-Y. Lee, M.~C. Vuran, and S.~Mohanty, ``{NeXt} generation/dynamic spectrum access/cognitive radio wireless networks: A survey,'' \emph{Computer Networks}, vol.~50, no.~13, pp. 2127--2159, 2006.

\bibitem{Manco22}
J.~Manco, I.~Dayoub, A.~Nafkha \emph{et~al.}, ``{Spectrum Sensing Using Software Defined Radio for Cognitive Radio Networks: A Survey},'' \emph{IEEE Access}, 2022.

\bibitem{Smith23}
P.~Smith, A.~Luong, S.~Sarkar \emph{et~al.}, ``{A Novel Software Defined Radio for Practical, Mobile Crowdsourced Spectrum Sensing},'' \emph{IEEE Trans. on Mob. Comp.}, 2023.

\bibitem{Cao24}
X.~Cao, B.~Yang, and K.~o. Wangand, ``{AI-Empowered Multiple Access for 6G: A Survey of Spectrum Sensing, Protocol Designs, and Optimizations},'' \emph{Proc. of the IEEE}, 2024.

\bibitem{Zhao19}
Y.~Zhao, S.~Luo, Z.~Yuan, and R.~Lin, ``{A New Spectrum Prediction Method for UAV Communications},'' in \emph{{2019 IEEE ICCC}}, 2019.

\bibitem{Cullen23}
A.~C. Cullen, B.~I.~P. Rubinstein, S.~Kandeepan, B.~Flower, and P.~H.~W. Leong, ``\BIBforeignlanguage{en}{Predicting dynamic spectrum allocation: a review covering simulation, modelling, and prediction},'' \emph{\BIBforeignlanguage{en}{Artif. Intell. Rev.}}, vol.~56, no.~10, pp. 10\,921--10\,959, Oct. 2023.

\bibitem{Zhao21Adhoc}
Z.~Zhao, M.~C. Vuran, B.~Zhou, M.~M. Lunar, Z.~Aref, D.~P. Young, W.~Humphrey, S.~Goddard, G.~Attebury, and B.~France, ``A city-wide experimental testbed for the next generation wireless networks,'' \emph{Ad Hoc Networks}, vol. 111, p. 102305, Feb. 2021.

\bibitem{OneLNK21}
\BIBentryALTinterwordspacing
M.~M.~R. Lunar, J.~Sun, J.~Wensowitch, M.~Fay, H.~B. Tulay, V.~S. S.~L. Karanam, B.~Qiu, D.~Nadig, G.~Attebury, H.~Yu, J.~Camp, C.~E. Koksal, D.~Pompili, B.~Ramamurthy, M.~Hashemi, E.~Ekici, and M.~C. Vuran, ``Onelnk: One link to rule them all: Web-based wireless experimentation for multi-vendor remotely accessible indoor/outdoor testbeds,'' in \emph{Proceedings of the 15th ACM Workshop on Wireless Network Testbeds, Experimental Evaluation \& CHaracterization}, ser. WiNTECH '21.\hskip 1em plus 0.5em minus 0.4em\relax New York, NY, USA: Association for Computing Machinery, 2021, p. 85–92. [Online]. Available: \url{https://doi.org/10.1145/3477086.3480835}
\BIBentrySTDinterwordspacing

\bibitem{Gemma24}
G.~Team, T.~Mesnard, C.~Hardin \emph{et~al.}, ``{Gemma: Open models based on gemini research and technology},'' \emph{arXiv:2403.08295}, 2024.

\bibitem{GPT19}
\BIBentryALTinterwordspacing
A.~Radford, K.~Narasimhan, T.~Salimans, and I.~Sutskever, ``{Language Models are Unsupervised Multitask Learners},'' 2019. [Online]. Available: \url{https://cdn.openai.com/better-language-models/language_models_are_unsupervised_multitask_learners.pdf}
\BIBentrySTDinterwordspacing

\bibitem{LLaMA23}
\BIBentryALTinterwordspacing
H.~Touvron, T.~Lavril, G.~Izacard \emph{et~al.}, ``{LLaMA: Open and Efficient Foundation Language Models},'' 2023. [Online]. Available: \url{https://arxiv.org/abs/2302.13971}
\BIBentrySTDinterwordspacing

\bibitem{Mistral23}
\BIBentryALTinterwordspacing
A.~Q. Jiang, A.~Sablayrolles, A.~Mensch \emph{et~al.}, ``{Mistral 7B},'' 2023. [Online]. Available: \url{https://arxiv.org/abs/2310.06825}
\BIBentrySTDinterwordspacing

\bibitem{Phi23}
\BIBentryALTinterwordspacing
S.~Gunasekar, Y.~Zhang, J.~Aneja \emph{et~al.}, ``{Textbooks Are All You Need},'' 2023. [Online]. Available: \url{https://arxiv.org/abs/2306.11644}
\BIBentrySTDinterwordspacing

\bibitem{Abd24}
M.~Abd~Elaziz, M.~A. Al-qaness, and A.~o. Dahou, ``{Evolution toward intelligent communications: Impact of deep learning applications on the future of 6G technology},'' \emph{Wiley Interdisciplinary Reviews}, 2024.

\bibitem{Matsumine24}
T.~Matsumine and H.~Ochiai, ``{Recent Advances in Deep Learning for Channel Coding: A Survey},'' \emph{IEEE Open Journal of the Comm. Society}, 2024.

\bibitem{Guo24}
J.~Guo, T.~Chen, S.~Jin \emph{et~al.}, ``Deep learning for joint channel estimation and feedback in massive mimo systems,'' \emph{Digital Communications and Networks}, vol.~10, no.~1, pp. 83--93, 2024.

\bibitem{Gao24}
S.~Gao, P.~Dong, Z.~Pan, and X.~You, ``{Lightweight deep learning based channel estimation for extremely large-scale massive MIMO systems},'' \emph{IEEE Transactions on Vehicular Technology}, vol.~73, no.~7, 2024.

\bibitem{Wang25}
D.~Wang, Z.~Wang, H.~Zhao \emph{et~al.}, ``{Secure Energy Efficiency for ARIS Networks With Deep Learning: Active Beamforming and Position Optimization},'' \emph{IEEE Trans. on Wireless Comm.}, 2025.

\bibitem{Kang24}
J.-M. Kang, ``{Deep Learning Enabled Multicast Beamforming With Movable Antenna Array},'' \emph{IEEE Wireless Comm. Letters}, 2024.

\bibitem{Shin25}
S.~Shin and M.~C. Vuran, ``{I Can't Believe It's Not Real: CV-MuSeNet: Complex-Valued Multi-Signal Segmentation},'' in \emph{{IEEE DySPAN}}, 2025.

\bibitem{Kumar24}
A.~Kumar, N.~Gaur, S.~Chakravarty \emph{et~al.}, ``{Analysis of spectrum sensing using deep learning algorithms: CNNs and RNNs},'' \emph{Ain Shams Engineering Journal}, 2024.

\bibitem{El24}
N.~El-haryqy, Z.~Madini, and Y.~Zouine, ``{A review of deep learning techniques for enhancing spectrum sensing and prediction in cognitive radio systems: approaches, datasets, and challenges},'' \emph{International Journal of Computers and Applications}, vol.~46, no.~12, 2024.

\bibitem{Subedi24LCN}
P.~Subedi, S.~Shin, and M.~C. Vuran, ``{Seek and Classify: End-to-end Joint Spectrum Segmentation and Classification for Multi-signal Wideband Spectrum Sensing},'' in \emph{{IEEE LCN'24}}, 2024.

\bibitem{Doke24}
K.~Doke, B.~Okoro, A.~Zare, and M.~Zheleva, ``{VIA: Establishing the link between spectrum sensor capabilities and data analytics performance},'' in \emph{{IEEE INFOCOM 2024}}.

\bibitem{Uvaydov21}
D.~Uvaydov, S.~D’Oro, F.~Restuccia \emph{et~al.}, ``{DeepSense: Fast Wideband Spectrum Sensing Through Real-Time In-the-Loop Deep Learning},'' in \emph{{IEEE INFOCOM 2021}}.

\bibitem{xing2022spectrum}
H.~Xing \emph{et~al.}, ``{Spectrum sensing in cognitive radio: A deep learning based model},'' \emph{Transactions on Emerging Telecommunications Technologies}, vol.~33, no.~1, p. e4388, 2022.

\bibitem{Lees19}
W.~M. Lees \emph{et~al.}, ``{Deep Learning Classification of 3.5-GHz Band Spectrograms With Applications to Spectrum Sensing},'' \emph{IEEE Transactions on Cognitive Communications and Networking}, 2019.

\bibitem{Zhang21}
W.~Zhang \emph{et~al.}, ``{Signal Detection and Classification in Shared Spectrum: A Deep Learning Approach},'' in \emph{{IEEE INFOCOM 2021}}.

\bibitem{Perenda21}
E.~Perenda \emph{et~al.}, ``{Learning the unknown: Improving modulation classification perf. in unseen scenarios},'' in \emph{{IEEE INFOCOM 2021}}.

\bibitem{Li20}
X.~Li, G.~Chen, Y.~Xu \emph{et~al.}, ``{Recovering Missing Values From Corrupted Historical Observations: Approaching the Limit of Predictability in Spectrum Prediction Tasks},'' \emph{IEEE Access}, 2020.

\bibitem{Zhang24}
W.~Zhang, Y.~Wang, X.~Chen \emph{et~al.}, ``{Spectrum Transformer: An Attention-Based Wideband Spectrum Detector},'' \emph{IEEE Trans. on Wireless Comm.}, 2024.

\bibitem{Guangliang25}
P.~Guangliang, L.~Jie, and L.~Minglei, ``{Multi-channel multi-step spectrum prediction using transformer and stacked Bi-LSTM},'' \emph{China Communications}, 2025.

\bibitem{ettus2025usrp_n310}
{Ettus Research (National Instruments)}, ``{USRP N310}: Networked software‑defined radio,'' Product page, Ettus Research, Jun. 2025, available at \url{https://tinyurl.com/usrpn310} (Accessed: 2025-06-11).

\bibitem{tensorflow15}
\BIBentryALTinterwordspacing
M.~Abadi, A.~Agarwal, P.~Barham \emph{et~al.}, ``{ {TensorFlow}}: Large-scale machine learning on heterogeneous systems,'' 2015, software available from tensorflow.org. [Online]. Available: \url{https://www.tensorflow.org/}
\BIBentrySTDinterwordspacing

\bibitem{Paszke2019pytorch}
A.~Paszke, S.~Gross, F.~Massa \emph{et~al.}, ``{Pytorch: An imperative style, high-performance deep learning library},'' \emph{Advances in neural information processing systems}, vol.~32, 2019.

\bibitem{oliveira2023trimmed}
R.~I. Oliveira and L.~Resende, ``{Trimmed sample means for robust uniform mean estimation and regression},'' \emph{arXiv:2302.06710}, 2023.

\bibitem{kocak4937461trimmed}
C.~Kocak, E.~Bas, and E.~Egrioglu, ``{Trimmed Mean Dendritic Neuron Model Artificial Neural Network for Time Series Forecasting in Case of Outliers},'' \emph{Available at SSRN 4937461}, 2024.

\bibitem{huggingface2025llama}
{Hugging Face \& EleutherAI}, ``Llama model configuration — transformers main branch,'' \href{https://github.com/huggingface/transformers/blob/main/src/transformers/models/llama/configuration_llama.py}{GitHub Repository}, 2025, accessed: 2025-06-14.

\bibitem{wolf2019huggingface}
T.~Wolf, L.~Debut, V.~Sanh \emph{et~al.}, ``{Huggingface's transformers: State-of-the-art natural language processing},'' \emph{arXiv:1910.03771}, 2019.

\bibitem{Garza2023}
A.~Garza and M.~Mergenthaler-Canseco, ``{TimeGPT-1: Foundation Model for Time Series Forecasting},'' \emph{arXiv}, 2023.

\bibitem{Kohen1960coefficient}
J.~Kohen, ``{A coefficient of agreement for nominal scale},'' \emph{Educ Psychol Meas}, vol.~20, pp. 37--46, 1960.

\end{thebibliography}

\end{document}